\documentclass[aps,prb,twocolumn,showpacs]{revtex4-1}

\usepackage{graphicx}% Include figure files
\usepackage{dcolumn}% Align table columns on decimal point
\usepackage{bm}% bold math
\usepackage[normalem]{ulem}
\usepackage{color}

\begin{document}

\title{$^{\bf 4}$He adsorbed outside a single carbon nanotube}  

\author{M.C.Gordillo}
\affiliation{Departamento de Sistemas F\'{\i}sicos, Qu\'{\i}micos 
y Naturales. Facultad de Ciencias Experimentales. Universidad Pablo de Olavide. 
Carretera de Utrera, km 1. 41013 Sevilla. Spain.}
\author{J. Boronat}
\affiliation{Departament de F\'\i sica i Enginyeria Nuclear,
Universitat Polit\`ecnica de Catalunya,
B4-B5 Campus Nord. 08034 Barcelona. Spain.}

\date{\today}

\begin{abstract}
The phase
diagrams  of $^4$He adsorbed on the external surfaces of single armchair
carbon  nanotubes  with radii in the range 3.42 -- 10.85 \AA \
are calculated using the diffusion Monte Carlo method.  For
nanotubes narrower than a (10,10) one,  the ground state
is an incommensurate solid similar to the one found   for
H$_2$ on the same substrates.   For wider nanotubes, the
phase with the minimum energy per  particle is a liquid layer. Curved
$\sqrt 3 \times \sqrt 3$ registered solids  similar to the ones found on
graphene and graphite were   unstable for all the tubes considered. 
\end{abstract}

% insert suggested PACS numbers in braces on next line
% insert suggested keywords - APS authors don't need to do this
%\keywords{}

\maketitle

\section{Introduction}

Until recently, the only experimental studies  about the adsorption
of quantum gases ($^4$He, H$_2$ and Ne) on  carbon nanotubes
were performed in tube associations called
bundles.~\cite{Dillon,talapatra,talapatra2,teizer,vilchesjltp1,vilchesjltp2,glyde} 
In these structures, the only places available for adsorption
are their  outer surfaces, in particular the grooves between two
adjacent tubes.~\cite{prl2008} Other possible
locations are the triangular interstices 
among three neighboring tubes in the bulk part of the bundle.
However, those places seem to be not
populated.~\cite{talapatra,talapatra2,prl2006}  The other option
is the adsorption on the inner surface of nanotubes,
but to do so one has to remove the caps
that close them at the ends.    To our knowledge, 
there is no experimental data on this possibility for quantum gases,
the only studies being up to now theoretical.~\cite{coleRMP,reatto,reattolow,prbr,prb07} 

A recent experimental work showed that it is
possible to isolate a single  carbon nanotube and adsorb Ar and Kr on its
outer surface.~\cite{vilchesscience}  The experimental
phase diagrams for these classical noble gases  were qualitatively similar
to those on graphite.   For instance, they include a $\sqrt 3 \times \sqrt
3$ phase for Ar on tubes whose diameters are in the range 1-3 nm. However,
recent quantum Monte Carlo  simulations~\cite{prb2011} 
of H$_2$ adsorbed on nanotubes with  diameters
between $\sim$ 7 and     22 \AA$ $ indicate that this structure was formed
only on the widest tube considered,  a (16,16) one in the standard
nomenclature.~\cite{reich}  Contrarily to flat surfaces,
mainly graphite, in which the phase diagrams of quantum gases are quite
well known from both theory and experiment, adsorption on curved surfaces
is much less studied. Apart from studies related to wetting transitions of
classical gases on cylinders and spheres,~\cite{classic1,classic2} the
possible new phases that curvature can produce in quantum gases are still
rather unknown~\cite{susana} and it is plausible to imagine that
experiments on isolated nanotubes can be exposed to quantum gases in a
nearby future.

In this work, we have used the diffusion Monte Carlo
(DMC) method  to obtain the  phase diagrams of $^4$He
adsorbed on isolated carbon nanotubes with radii ranging between 3.42 and
10.85 \AA. These correspond to armchair nanotubes between (5,5)
and (16,16) in the standard nomenclature. Our main aim is to compare our
results with both the ones for H$_2$ on the same 
surface~\cite{prb2011} and those for $^4$He on graphene,~\cite{prl2009}
a flat surface that can be considered  
a tube of infinite  radius. 

The rest of the paper is structured as follows. In the next section,
we  describe the DMC algorithm used  to solve the corresponding
Schr\"odinger equations from which we   inferred the corresponding phase
diagrams. The phase diagrams themselves will be shown in Section III, and
the paper will close  by summarizing our  main conclusions in  
Section IV.   

\section{Method}
The diffusion Monte Carlo method allows for
solving the many-body Schr\"odinger equation exactly if the constituents
of the system are bosons,~\cite{boro94} as it is the case of 
$^4$He atoms. Since it is a stochastic method,  the
results are always produced with some  statistical uncertainties.
The  different carbon nanotubes were considered as rigid cylinders in which
the individual carbon positions were kept constant throughout the
simulation. The interaction between any of those carbons and each helium
atom was of the Lennard-Jones type with parameters taken  
from Ref. \onlinecite{cole}, i.e., the carbon-helium
interactions act as an external potential to be included
in the Schr\"odinger equation to solve.    The helium-helium interparticle
interaction was taken to be a standard  Aziz
potential.~\cite{aziz}

DMC solves the $N$-body Schr\"odinger equation in imaginary time. The
walkers (sets of $3N$ coordinates) representing the system evolve in time
through a diffusion process and an energy selection rule named branching. 
In order to reduce the statistical variance, the free diffusion movement is
corrected by a drift term that enhances the sampling in regions where the
exact wave function is reasonably expected to be large. Technically, this
importance sampling is implemented by introducing a guiding wave function
which avoids the sampling inside the cores of the interaction and fixes the
phase of the system under study. In particular, for an adsorbed liquid
layer we use
\begin{equation}
\Phi({\bf r}_1, {\bf r}_2, \ldots, {\bf r}_N) = \prod_{i<j} \exp \left[-\frac{1}{2} 
\left(\frac{b_{\rm He-He}}{r_{ij}} \right)^5 \right] \prod_i
\Psi(r_i) \ ,
\label{trial1}
\end{equation} 
where ${\bf r}_1, {\bf r}_2, \ldots, {\bf r}_N$ are the $^4$He positions. 
The first part of the
product in Eq. (\ref{trial1}) is a  Jastrow function that depends on the 
distances $r_{ij}$
between the atoms.    There, $b_{\rm He-He}$, is a variational parameter
fixed at 3.07 \AA.~\cite{prl2009}   $\Psi(r_i)$ is the same one-body
function than the one used in Ref. \onlinecite{prl2009}, whose role is to
radially confine the moving particles around the minimum of the collective sum of
all C-He interactions. To do so, it has a maximum   at a distance of 
$\sim 2.87$ \AA \  from the surface of the different tubes and decreases
monotonously both for larger and smaller separations. 

When the cylinder is coated with a solid layer,
we introduce in the trial wave function a factor, 
\begin{equation} 
\prod_i \exp \left\{ -c \left[
(x_i-x_{\text{site}})^2+ (y_i - y_{\text{site}})^2 +  (z_i -
z_{\text{site}})^2 \right] \right\}  \ , 
\label{trialsol} 
\end{equation}
that multiplies the Jastrow part (Eq. \ref{trial1}). With this
approach, 
each helium atom is mainly sampled in the vicinity of its
associated  crystallographic position
$x_{\text{site}},y_{\text{site}},z_{\text{site}}$.  The value of the $c$
parameter for the different solid phases were the same as 
the ones used  for the same phases in a previous simulation on
graphene.~\cite{prl2009}

\section{Results}

The nanotubes considered in this work, 
together with their respective radii (distance from the center of the tube to the 
carbon surface) and the energy per particle in the infinite dilution limit,
$e_0$,  are 
shown in Table \ref{table1}. Those energy values were obtained from simulations 
including a single helium atom on a tube 70 \AA \  long.  What we observe is that, 
in general $e_0$ increases with the tube radius, but it is far from the value of 
a flat surface. In that Table, $r_0$ is the most probable radial distance   
of a single helium atom to the center of each tube. In the remaining of the paper, 
all the adsorbate densities will be calculated using that distance 
as the radius of the helium cylinder that coats the carbon nanotube. 

There is no simple explanation for the evolution of $e_0$  
as a function of the tube radius beyond a general 
increase when it increases. For instance, $e_0$ is slightly lower  
for the (14,14) tube than for the (16,16) one. To try to understand this
effect and its connection with the curvature of the adsorbing surface,
we calculated the difference in the potential energy 
felt by a helium particle located 2.87 \AA \  above the center of one of the
hexagons that conform the tube ($c$) and for an 
atom located above the center of a C-C bond ($s$) at the same height. 
This difference is an estimation of 
the corrugation of the potential surface. In graphene,
a flat structure, those values are $V_c=-175.02$ and  $V_s=-167.49 $ K. 
This means a difference of 7.53 K. That is a measure of the energy barrier that
$^4$He atoms have to jump in order to explore the graphene surface. 
The corresponding
average potential energy from a full DMC calculation of a single helium on that 
structure is -151.43 $\pm$ 0.05 K, what indicates that the particle 
explores also less favorable positions. 
The respective pairs of potential
energies for the tubes under consideration are $V_c=-154.35$ and
$V_s=-147.57$ K (a difference of 6.79 K) for the (14,14) tube, 
to be compared to $V_c=-156.32$ and $V_s=-148.19$ K (8.13 K) for the (16,16) one.
The DMC  potential energy averages are -125.7 $\pm$ 0.1 and -125.6  $\pm$ 0.1 K, 
respectively, i.e., virtually identical. This means that simply 
summing up the He-carbon interactions in a reduced number of positions  
is not a precise guide neither for $e_0$ nor for the 
average potential interaction. The corrugation effect in a single
carbon hexagon is similar for a planar and for a curved environment but the
potential energy and $e_0$ are smaller (in absolute value) than in planar graphene 
because the curved adsorbent is globally less attractive.

%The values of Table \ref{table1} consider
%also the kinetic energy terms, in our case 20.5 $\pm$ 0.1 K and 20.6 $\pm$ 0.1 K for the (14,14) and
%(16,16) tubes, respectively. In general, the kinetic energy increases when the
%potential energy decreases. For instance, the same value a single particle 
%adsorbed on graphene is 21.72 $\pm$ 0.04 K, to be compared to 19.69 $\pm$ 0.05 K, the same value for 
%a (5,5) tube. The corresponding potential energy is -114.45 $\pm$ 0.05 K.    

\begin{table}
\caption{The set of armchair carbon nanotubes considered in this work,
their tube radii ($r_t$) together with the helium adsorption energy ($e_0$) in
the infinite dilution limit. Also included for comparison is the same value for
graphene, taken from Ref. \onlinecite{prl2009}. $r_0$ is the most probable
distance of a single $^4$He atom to the center of the tube.  
}
\begin{tabular}{cccc} \hline
Tube & $r_t$ (\AA) &  $r_0$ (\AA)   & $e_0$ (K) \\ \hline
(5,5) & 3.42 &  6.26  & -94.76 $\pm$ 0.06  \\
(6,6) & 4.10 &  6.93  & -97.4  $\pm$ 0.2  \\
(8,8) & 5.45 &  8.26  & -100.94 $\pm$ 0.09 \\
(10,10) & 6.80  &  9.65 & -103.29 $\pm$  0.08 \\
(12,12) & 8.14 & 11.01  & -104.72  $\pm$ 0.04 \\
(14,14) & 9.49 &  12.37 & -105.22 $\pm$  0.07  \\
(16,16) & 10.85 & 13.70 & -104.98 $\pm$  0.06  \\
graphene (Ref. \onlinecite{prl2009}) & $\infty$ & -- & -141.64 $\pm$ 0.01 \\ 
 \hline
\end{tabular}
\label{table1}
\end{table}

To study the possible helium phases on each carbon cylinder,  we follow
closely a previous study of H$_2$ adsorbed on the same systems (Ref.
\onlinecite{prb2011}).  Liquid phases, i.e., helium
arrangements described by Eq. \ref{trial1}, and curved counterparts of the
registered $\sqrt 3 \times \sqrt 3$ phases have been 
considered.~\cite{vilchesscience}   
Structures with  helium atoms regularly
distributed on circumferences (or rows) whose defining planes were
perpendicular to the main axis ($z$) of the cylinders were also studied. Those
solids are defined by the number   of absorbed atoms on each circle and are
named accordingly. For instance, if we have ten atoms on the outside of the
circle, the phase is termed a ten-in-a-row solid.  The densities were
determined by the distances between neighboring  circumferences along the
$z$ axis.  Those rows were  rotated  half the distance between helium atoms
with respect to each other to maximize the number of closest neighbors
for each helium atom. For completeness,  the wrapped up equivalents to
the $3/7$ and $2/5$ registered structures proposed by  Greywall~\cite{grey2}
for graphite were also taken into account. 

\begin{figure}[b]
\begin{center}
\includegraphics[width=7.5cm]{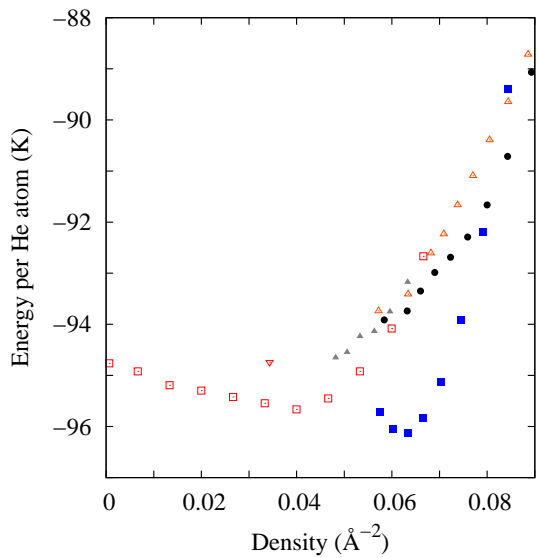}
\caption{(Color online) Energy per helium atom for different phases
adsorbed on a (5,5) tube. Open squares, liquid phase; full squares, a
five-in-a-row atom solid incommensurate phase; full circles, six-in-a-row
arrangement; open  up triangles, incommensurate solid  with seven atoms
per row;  up full triangles, solid of the same type with four atoms per
circumference.  The open down triangle indicates the result for a
curved $\sqrt 3 \times \sqrt 3$ structure.  }
\label{energy1}
\end{center}
\end{figure}

Fig ~\ref{energy1} shows the phase diagram of $^4$He adsorbed on the
surface of a (5,5) tube. We observe that, as in the case of H$_2$ adsorbed
on the same cylinder,~\cite{prb2011} the ground state is not a liquid
(open squares), but a structure corresponding to a five-in-a-row solid
(full squares). The minimum energy per particle is $-96.10 \pm 0.03$ K with
a density  $\rho = 0.0620 \pm 0.0002$ \AA$^{-2}$, as can be seen in Table
\ref{table2}. These results,  as the ones corresponding to incommensurate
solids on other tubes, were extracted from a third-order polynomial fit to
the simulation results around the point of the minimum energy; in this particular case, to
the six points corresponding to the lowest densities of the set of full  squares in Fig ~\ref{energy1}. 
The error bars in Table
\ref{table2} were also obtained from  the
least-squares fit. $\rho = 0.0620$ \AA$^{-2}$ is the lowest  density for which a 
solid is stable; below
it, the system will form a patch of that solid  and leave the rest of
surface of the tube empty. On the other hand,  when the amount of adsorbed
helium increases, the energy per particle of this five-in-a-row structure
is above the corresponding to a six-in-a-row incommensurate solid (full
circles in Fig.~\ref{energy1}). This means that the system should undergo
a first-order phase transition between those arrangements.   By means of a
double-tangent Maxwell construction considering both the full squares and full circles
in Fig.~\ref{energy1}, we found that  the transition region
is between $0.0752 \pm 0.0002$ \AA$^{-2}$ (upper density limit for a
five-in-a-row structure), and $0.0845 \pm 0.0002$ \AA$^{-2}$ (lower density
for which a six-in-a-row structure is stable).  
Their respective energies
per particle are $-93.67 \pm 0.01$  and $-90.63 \pm 0.01$ K. 
To estimate these energies and densities, we used the simulation results and 
interpolated
between them a set of third order polynomial splines. This means that no particular
polynomial fit for any of the equation of state (represented here by the complete 
set of simulation points) was used.  
None of the
other arrangements considered (liquid, registered $\sqrt 3 \times \sqrt 3$
solid, or other commensurate structures) are stable, since their energies
per particle are larger than  the corresponding to  any of the above
mentioned structures (see Fig.~\ref{energy1} and Table \ref{table2}). For
the registered solids whose data are given in Table \ref{table2}, the
densities are given as exact values, since a pure commensurate structure
forms only at a characteristic density. The error bars of
the energies
per particle are the statistical uncertainties   derived from their
respective  diffusion Monte Carlo calculations. 

The phase diagram for a
(6,6) tube is similar to the already described for the (5,5) one: the
ground state is a six-in-a-row structure whose minimum density obtained from the same
type of third order polynomial fit to the simulation results around the minimum of energy, is 
$0.0658 \pm 0.0005$ \AA$^{-2}$. Upon a density increase,   
this solid structure suffers a
first-order phase transition at $\rho = 0.0836 \pm 0.0002$ \AA$^{-2}$
(energy per particle, $-95.68 \pm 0.03$ K) to a seven-in-a-row solid  with
$\rho = 0.0843 \pm 0.0003$ \AA$^{-2}$ (energy per particle, $-91.78 \pm
0.02$ K).       That neither of the other commensurate or liquid phases are
stable can be checked in Table \ref{table2}: all of them have energies per
particle greater that the one for the ground state.  

\begin{figure}[b]
\begin{center}
\includegraphics[width=7.5cm]{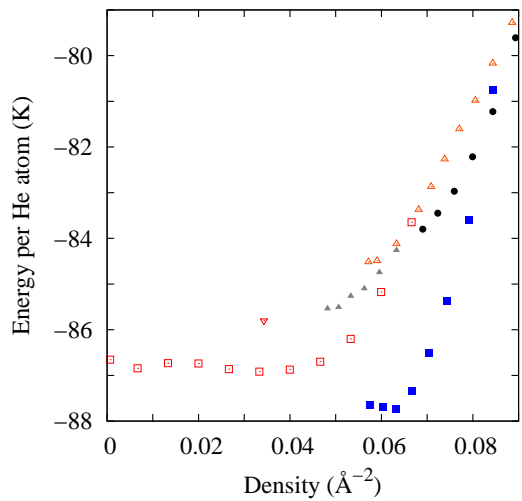}
\caption{(Color online) Same that in the previous figure, but using as a 
C-He potential the anisotropic expression of 
Carlos and Cole (Ref. \onlinecite{carlosandcole}). 
}
\label{energy2}
\end{center}
\end{figure}

\squeezetable
\begin{table*}
\caption{Energies per particle ($e_b$) and equilibrium helium densities
($\rho$) for nanotubes in which the  
ground state is an incommensurate solid.   
The error bars are given in parenthesis and affect to the last figure shown. 
The commensurate solid arrangements are represented by a single density without error bars. 
The results labeled with a $^*$ were obtained using the anisotropic potential of 
Ref. \onlinecite{carlosandcole} 
instead of the isotropic interaction employed in the other cases. 
}
\begin{tabular}{ccccccccccc} \hline
Phase   & & liquid &  & 2/5 & & 3/7 & & $\sqrt{3} \times \sqrt{3}$  & & incommensurate solid \\ \hline
Tube    & $\rho$ (\AA$^{-2}$)  & $e_b$ (K) & $\rho$  (\AA$^{-2}$)  & $e_b$ (K) & 
       $\rho$  (\AA$^{-2}$)  & $e_b$ (K) & $\rho$  (\AA$^{-2}$)  & $e_b$ (K) & 
       $\rho$  (\AA$^{-2}$)  & $e_b$ (K)  
      \\ \hline
(5,5) & 0.0364(7) & -95.66(3) & 0.0410 & -95.07(8) & 0.0439 & -94.63(8) &  0.0341 & -94.74(3) & 0.0620(2) & -96.10(3)  \\
(5,5)$^*$ & 0.0364(5) & -86.94(2) & 0.0410 & -86.70(8)  &0.0439 & -85.93(7) &  0.0341 & -85.81(3) & 0.0598(2) & -87.76(5)  \\
(6,6) & 0.0356(2) & -98.65(6) & 0.0444 & -97.81(6) & 0.0475 & -97.07(7) &  0.0369 & -97.01(4) & 0.0658(5) & -99.05(4)   \\
(8,8) & 0.0358(1) & -102.02(2) & 0.0495 & -101.49(4) & 0.0529 & -100.56(4) &  0.0412 & -101.00(4) & 0.0744(9) & -102.50(2) \\
(10,10)& 0.0323(9)& -104.20(2) & 0.0534 & -103.74(5) & 0.0573 & -102.78(3) &  0.0445 & -103.36(3) & 0.0730(7) &  -104.15(4)  \\
(10,10)$^*$ & 0.0373(1)& -98.31(2) & 0.0534 & -98.12(5) & 0.0573 & -96.65(5) &  0.0445 & -97.93(5) & 0.0783(4) &  -98.88(4)  \\
 \hline
\end{tabular}
\label{table2}
\end{table*}

To obtain the data displayed in Fig.~\ref{energy1}, we have used the
isotropic C-He interaction of Stan and Cole  (Ref. \onlinecite{cole}). This
was the interaction used in previous studies of helium adsorbed on carbon
nanotubes (see for instance Refs. \onlinecite{coleRMP,
prbr,prb07,yolow}). However, there is another possibility: an anisotropic
form of the same interaction proposed by Carlos and
Cole.~\cite{carlosandcole} The DMC results with this latter
interaction are displayed in Fig.~\ref{energy2} and Table \ref{table2}.
From them, one can see that the ground state is still a
five-in-a-row solid whose minimum stable density is $\sim 3.5$ \% lower
than the one for the isotropic potential. The energies per particle
in the anisotropic case are $\sim 8.7$ K larger due to a
slight increase of its potential well depth with respect to the isotropic
model.  However, there is no way to discriminate between the
two sets of parameters since, to our knowledge, there is not available 
experimental information to compare our simulations to.  Our results
show that with the anisotropic interaction
there is also a first-order phase transition at $\rho = 0.0787 \pm
0.0001$ \AA$^{-2}$ (energy per particle, $-83.77 \pm 0.01$ K) to a
six-in-row solid whose lowest  density is $0.0883 \pm 0.0001$ \AA$^{-2}$
and the energy per particle is $-80.07 \pm 0.02$ K. This
means that the transition region is shifted to larger densities. We can
also see in Table \ref{table2} that the anisotropic interaction favors the
incommensurate solid with respect to the liquid: the energy per
particle difference  between the ground state and the liquid minimum is
$0.44 \pm 0.04$ K for the isotropic potential and $0.82 \pm 0.05$ K for the
anisotropic one.

\begin{figure}[b]
\begin{center}
\includegraphics[width=7.5cm]{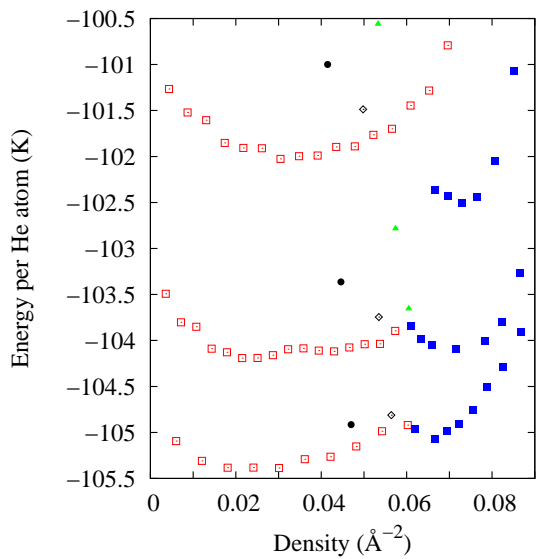}
\caption{(Color online) Energy per particle as a function of density for liquid (open squares),   
incommensurate (full squares), $\sqrt 3 \times \sqrt 3$ (full circle), 3/7 (full triangles) and
2/5 (open diamonds) phases. The upper set of symbols corresponds to a (8,8) tube, the middle set
to a (10,10) tube and the lowest to a (12,12) cylinder.   
}
\label{energy3}
\end{center}
\end{figure}

The phase diagram of the (8,8) tube is qualitatively similar to those of
the thinner tubes already  described. Its ground state is an incommensurate
solid of the type $n$-in-a-row, were $n$ is the nanotube index  ($n= 8$).
The  main difference is that in the density range considered (up to $\sim
0.10$ \AA$^{-2}$),  there is only one stable
incommensurate structure, i.e., there is no first-order transition
between  two incommensurate solids. Both in the upper part of
Fig.~\ref{energy3} and in Table \ref{table2}, it can be seen  that
neither the liquid phase nor any of the registered solids are stable with
respect to the incommensurate structure.  For the
rest of nanotubes explored in this work, all the incommensurate solids
considered will be of the type $n$-in-a-row, where $n$ is the nanotube
index.

In Fig.~\ref{energy3}, we examine how the phase diagrams evolved when the
tube radii increased. There, from top to bottom we show
the energy per particle of the different adsorbed phases on
the (8,8), (10,10) and (12,12) tubes.    As indicated above,
for the (8,8) nanotube the ground state 
corresponds to an incommensurate solid. By increasing even more
the tube radius we observe a progressive tendency towards stability of the
liquid phase.  In particular, our results reported in 
Table \ref{table2} indicate  that the liquid phase absorbed on a (10,10)
tube is essentially as stable as a ten-in-a-row solid, since their energies
per particle are similar within their respective error bars. This is not
true for the  (12,12) cylinder, in which the ground state is clearly a
liquid. A third-order polynomial fit to
the liquid energies  indicates that its equilibrium density  is
$0.0241 \pm 0.0009$ \AA$^{-2}$ and its  energy per
helium atom at this density  $-105.40 \pm 0.02$ K. A
double-tangent  Maxwell construction allows us to say that upon a
density increase,  a liquid at
density $\rho = 0.0396 \pm 0.002$ \AA$^{-2}$ undergoes a first-order
transition to a twelve-in-a-row solid   whose lowest stable density
is $0.0667 \pm 0.003$ \AA$^{-2}$. The energies corresponding to the
$2/5$ and  $3/7$ commensurate structures, displayed as  open
diamonds and full triangles in Fig.~\ref{energy3}, are  bigger than those
of the liquids with the same density, and therefore, unstable structures.
We can also see that the curved equivalents to the $\sqrt 3 \times \sqrt 3$
registered solid, whose energies  per particle are displayed as full
circles, are not stable either.   The use of an anisotropic C-He potential
in the (10,10) nanotube has basically the same effect than in the
(5,5) one: it favors the stability of the incommensurate solid over the
liquid, that now it is $0.57 \pm 0.04$ K less stable than a ten-in-a-row
structure.   

\begin{figure}[b]
\begin{center}
\includegraphics[width=7.5cm]{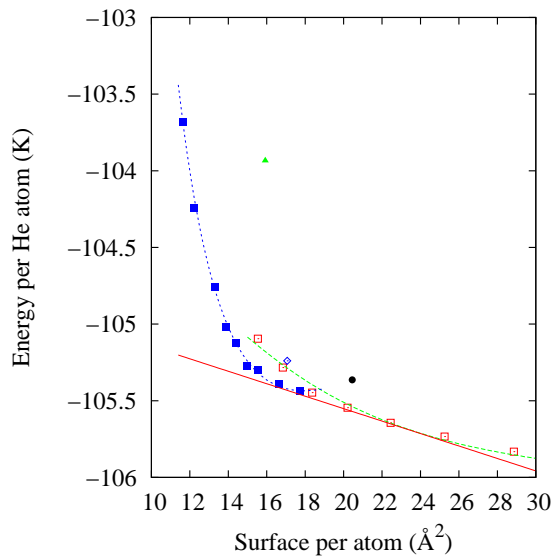}
\caption{(Color online)  Double-tangent Maxwell construction
solid line) used to obtain the limits of the 
transition zone between a liquid structure (dotted line, full squares), and an 
incommensurate one
(dashed line, open squares). The meaning of the other symbols is the same as in 
Fig.~\ref{energy3}.
}
\label{energy4}
\end{center}
\end{figure}

The other two cylinders considered, [(14,14) and (16,16)], have phase
diagrams similar to that of the  (12,12) tube. Their respective ground
states are liquids 
with equilibrium densities and energies at equilibrium
reported  in Table \ref{table3}. The double-tangent Maxwell
construction used in the (14,14) tube to obtain the limits of the
transition zone between the liquid and the corresponding fourteen-in-a-row
solid is shown in Fig.~\ref{energy4}. The energy  and density limits of the
transition zones for those three tubes are given in Table \ref{table3}. 
All the commensurate solids ($2/5$,$3/7$, and $\sqrt 3 \times \sqrt 3$) are
unstable with respect both to the liquid and to the incommensurate solids
in all the cylinders considered. This can be seen in  Fig.~\ref{energy4}
for the (14,14) nanotube, were their energies per particle are
represented as an open diamond, a full triangle and a full circle,
respectively. From Table \ref{table3} we are also able to deduce that the
equilibrium densities for the liquid phases (whose average density is
$\sim 0.023$ \AA$^{-2}$) are much lower than their metastable counterpart 
in graphene~\cite{prl2009} (0.044 \AA$^{-2}$) and the same happens to the
lowest stability limit of an incommensurate phase: an average of $\sim
0.064$ \AA$^{-2}$  instead of the  value 0.08 \AA$^{-2}$ of Ref.
\onlinecite{prl2009}.     

\squeezetable
\begin{table*}
\caption{Energies per particle ($e_g$) and equilibrium 
$^4$He densities
($\rho_g$) for tubes in which the ground state is a liquid.   
As before, the error bars are given in parenthesis and affect to the last figure shown. 
The upper stability limits for the liquids ($\rho_u$), and the lower
densities for which the incommensurate solids are stable ($\rho_l$) are also given, together
with their corresponding energies per particle. 
}
\begin{tabular}{ccccccc} \hline
Tube    & $\rho_g$ (\AA$^{-2}$)  & $e_g$ (K) & $\rho_u$  (\AA$^{-2}$)  & $e_u$ (K) & 
       $\rho_l$  (\AA$^{-2}$)  & $e_l$ (K)  
      \\ \hline
(12,12) & 0.0241(9) & -105.40(2) & 0.040(1) & -105.28(1) & 0.067(1) & -105.05(2)  \\
(14,14) & 0.0189(9) & -106.01(2) & 0.043(1) & -105.70(2) & 0.060(1) & -105.42(5)   \\
(16,16) & 0.0260(1) & -105.85(2) & 0.054(1) & -105.44(5) & 0.066(1) & -105.15(4)  \\
 \hline
\end{tabular}
\label{table3}
\end{table*}

\section{Conclusions}

Diffusion Monte Carlo calculations has allowed us to study the phase 
diagrams of
$^4$He adsorbed on carbon nanotubes of different radius. 
In principle, the ground state of this system should be a curved counterpart
of the $\sqrt 3 \times \sqrt 3$ structure found in graphite and graphene. 
This is what have 
been experimentally observed for heavier noble gases~\cite{vilchesscience} adsorbed 
on wider tubes. 
In fact, any armchair nanotube ($n$,$n$) fulfills the necessary prescription for 
the formation of a seamless registered solid on it, i.e., its  
indexes ($n$,$m$) follow the restriction $(n-m)/3 = k$,  $k$ being 
an integer.   
However, as it can be inferred from the results of Table \ref{table2}, 
the ground state of $^4$He on those tubes is either a liquid or an incommensurate
solid, not a $\sqrt 3 \times \sqrt 3$ arrangement.    
A comparison to the case of H$_2$ on the same substrates,~\cite{prb2011} in
which the $\sqrt 3 \times \sqrt 3$ solid is the ground state for a (16,16) tube, 
indicates that, at least for the widest tubes considered, the
weakness of the helium-helium  and helium-carbon interactions are responsible 
for the lack of stability of that (or any other) commensurate structure. 

Another relevant feature we have found is that for tubes whose radius is lower
than $\sim 7$ \AA,  the ground state is an incommensurate solid, whereas
from the (12,12) tube up, the minimum  energy structure 
corresponds to a liquid. This is probably a consequence  of the above mentioned 
weakness in the helium interactions combined with the lower density of the
curved $\sqrt 3 \times \sqrt 3$ phase in comparison to planar graphene. 
Our results for the $^4$He phase diagrams show a significant dependence on
the curvature of the different tubes. 
As Table \ref{table1} shows, the energy in
the infinite  dilution limit for a (5,5) tube is $\sim 10$ K larger than
the same property for a (12,12) cylinder. This means that to obtain 
an energy per particle closer to the corresponding to a flat structure,
one would
have to increase more the number of helium neighbors of each 
adsorbed atom for a thinner tube than for a wider one. This leads to high
densities and the formation of curved solids.
These structures 
are incommensurate rather than registered because the necessary helium densities
are higher than the corresponding to a $\sqrt 3 \times \sqrt 3$ (or any other, 
see Table \ref{table2}) commensurate structure. 
For instance, 
in the (8,8) tube the density of a $\sqrt 3 \times \sqrt 3$ curved structure is
0.0412 \AA$^{-2}$ (Table \ref{table2}), to be compared to 0.0636 \AA$^{-2}$ for 
the same arrangement on graphite and to 0.0744  \AA$^{-2}$ of the incommensurate
solid on the same cylinder. However, when the diameter of the tube grows 
the  curvature of the substrate is reduced 
and arrangements with lower densities, i.e., liquids, become stable. 

\begin{acknowledgments}
We acknowledge partial financial support from the Junta de
de Andaluc\'{\i}a Group PAI-205, Grant No. FQM-5985, MICINN 
(Spain) Grants No. FIS2010-18356 and FIS2011-25275, and Generalitat de Catalunya
Grant 2009SGR-1003.
\end{acknowledgments}

\end{document}